\documentclass[12pt]{iopart}
\usepackage{graphicx}
\usepackage{epstopdf}
\usepackage{dcolumn}
\usepackage{bm}
\usepackage{color} 
\usepackage{CJK}
\usepackage{stfloats}

\def\la{{\langle}}
\def\ra{{\rangle}}

\newcommand{\beq}{\begin{equation}}
\newcommand{\eeq}{\end{equation}}
\newcommand{\beqa}{\begin{eqnarray}}
\newcommand{\eeqa}{\end{eqnarray}}

\begin{document}

\title[Optimal transport of two ions under slow spring-constant drifts]{Optimal transport of two ions under slow spring-constant drifts}

\author{Xiao-Jing Lu$^{1,2}$, Mikel Palmero$^2$, Andreas Ruschhaupt$^3$, Xi Chen$^1$ and Juan Gonzalo Muga$^{1,2}$}
\address{$^1$ Department of Physics, Shanghai University, 200444
Shanghai, People's Republic of China}
\address{$^2$ Departamento de Qu\'{\i}mica F\'{\i}sica, UPV/EHU, Apdo.
644, 48080 Bilbao, Spain}
\address{$^3$ Department of Physics, University College Cork, Cork, Ireland}

%

%
\begin{abstract}
We investigate the  effect of slow spring-constant drifts of the trap used 
to shuttle two ions of different mass. 
We design transport protocols to suppress or mitigate the final excitation energy by
applying invariant-based inverse engineering, perturbation theory,  and
a harmonic dynamical normal-mode approximation. A simple, explicit   
trigonometric protocol for the trap trajectory is found to be robust with respect to the 
spring-constant drifts.      
\end{abstract}
\maketitle
%
%
%

\section{Introduction}
A possible
scalable architecture for quantum information processing
relies on shuttling small numbers  of trapped ions among storing and processing sites
in multi-electrode configurations \cite{Wineland2002,Rowe,Wineland2006,Roos,Monroe}. Transport of two ions
of different species is particularly relevant as one of them may be used for cooling
and the other one to encode the qubit \cite{2dif}. 
Diabatic transport of two equal ions has been recently realized  \cite{Bowler,Schmidt}. 
It was recognized \cite{Bowler} that different masses would require special consideration since all modes
may be excited by the transport.  
On the theory side, equal-mass two-ion transport has been studied in \cite{Mikel}
to design fast protocols without final excitation by invariant-based inverse engineering,
whereas the design of fast transport protocols of two ions with different mass was tackled in \cite{Mikel2}
using a harmonic approximation in normal mode coordinates that is accurate up to very short transport times,
of the order of a few oscillations of the ions.

The transport protocols are subjected to noise and perturbations. In current experiments, the errors in the spring constant due to slow drifts of imperfect calibration
are likely to dominate others. This means that the spring constant for each run of
the experiment stays constant, but it may change from run to run, differing from the ideal value used to set the protocol.
The effect of these errors was studied in \cite{Mainz} for single
ion-transport.
For two ions of equal mass, the normal mode coordinates  become proportional to center-of-mass (CM) and relative coordinates and are exactly decoupled.
In that case, only the center of mass can be excited by the motion of a harmonic trap \cite{Mikel}, so that the results and techniques in \cite{Mainz} (valid 
for one ion or a decoupled center of mass motion) are directly applicable.
For unequal masses though, this decoupling of coordinates does not hold
so that a different approach is needed.

In this paper we investigate the effect of spring-constant perturbations on the transport of two ions of different
mass within a harmonic approximation in dynamically defined  normal-mode coordinates \cite{Mikel2}, and apply invariant-based inverse engineering combined with perturbation theory in the relative error parameter to design transport protocols that suppress or mitigate the final excitation energy.
In Sec. \ref{invariant} and \ref{normal mode} we briefly introduce the invariant-based inverse engineering method and the dynamical
normal modes; in Sec. \ref{spring constant error} we design protocols that suppress effectively the excitation energy up to very
small shuttling times, of interest for current quantum information processing applications.
\section{Invariant-based engineering method}\label{invariant}
In this section, we provide a brief review of invariant-based engineering for shuttling
one ion \cite{Erik1}. As the Hamiltonian is quadratic, the structure and properties of dynamical invariants and propagators 
are known   
\cite{ManDodo, other} and may be used to design the trap motion.   
The harmonic transport of one ion
is described by the effective 1D Hamiltonian
\beq
\label{Hamiltonian}
\hat{H}_0 (t)= \frac{\hat{p}^2}{2 m} +
\frac{1}{2}m \omega^2 [\hat{q}-Q_0(t)]^2,
\eeq
where $\hat{q}$ and $\hat{p}$ are the position and momentum
operators, $\omega/(2\pi)$ is the  frequency of the
trap, and $Q_0(t)$ the position of its moving center. The corresponding quadratic-in-momentum
Lewis-Riesenfeld invariant \cite{LR,LL,DL} is given  (up
to an arbitrary multiplicative constant) by \cite{Erik1}
\beq
\label{inva}
\hat{I} (t) = \frac{1}{2m}[\hat{p}-m\dot{\alpha}_c(t)]^2 +\frac{1}{2}m \omega^2 [ \hat{q}-\alpha_c (t)]^2,
\eeq
where the dot represents a time derivative, and the function $\alpha_c (t)$ must satisfy the auxiliary equation
\beq
\label{classical}
\ddot{\alpha}_c+\omega^2(\alpha_c-Q_0)=0,
\eeq
so that the invariant condition holds,
\beq
\frac{d \hat{I}(t)}{d t} \equiv \frac{\partial \hat{I}(t)}{ \partial t} +\frac{1}{i \hbar} [\hat{I}(t), \hat{H}_0(t)] =0.
\eeq
The expectation value of $\hat{I}(t)$ remains constant
for solutions of the time-dependent Schr\"odinger equation $i \hbar
\partial_t\Psi ({q},t) = \hat{H}_0(t) \Psi ({q},t)$. The solutions can be expressed in terms of independent ``transport modes''
$\Psi({q},t) = \sum_n c_n \Phi({q},t;n),$
where $\Phi({q},t;n)=e^{i\theta(n)} \phi({q},t;n)$,
$n=0,1,...$ is the mode index;
$c_n$ are time-independent coefficients;  and $\phi({q},t;n)$ are the
orthogonal eigenvectors of the invariant $\hat{I}(t)$ satisfying $\hat{I}(t)\phi({q},t;n)= \lambda(n)\phi({q},t;n)$,
with real time-independent eigenvalues $\lambda(n)$.
Finally, the Lewis-Riesenfeld phase is
\beq
\label{LRphase}
\theta(t;n) = \frac{1}{\hbar} \int_0^t \Big\langle \phi (t';n) \Big|
i \hbar \frac{\partial }{ \partial t'} - \hat{H}_0(t') \Big| \phi (t';n)  \Big\rangle d t'.
\eeq
For the harmonic trap \cite{Erik1},
\beq
\label{psin}
\phi({q},t;n) = \exp{\left(i \frac{ m\dot{\alpha}_c
{q}}{\hbar}\right)}\phi^{(0)}({q}-\alpha_c;n),
\eeq
where $\phi^{(0)}({q};n)$ are the eigenstates of Eq. (\ref{Hamiltonian}) for $Q_0(t)=0$.
Note that in harmonic transport $\alpha_c$ is the center
of the transport modes which obeys the classical Newton equation
(\ref{classical}). 

To transport the ion between $0$ and $d$ in a time $T$, the trajectory $Q_0$ should satisfy
\beq
\label{Q0}
Q_0(0)=0,~~Q_0(T)=d.
\eeq
The inverse engineering strategy is to design the invariant first, via $\alpha_c(t)$,
and then get $Q_0(t)$ from the Newton equation (\ref{classical}).
To guarantee the commutativity  of $\hat{I}(t)$ and
$\hat{H}_0(t)$ at initial time $t=0$ and final time $t=T$ (which implies the
shuttling from initial to final
trap eigenstates without final excitation), and the continuity of trap motion,
the designed $\alpha_c(t)$ should satisfy the boundary conditions \cite{Erik1}
\beqa
\alpha_c(0)=0,~~ \alpha_c(T)=d,
\nonumber\\
\dot{\alpha}_c(0)=0,~~\dot{\alpha}_c(T)=0,
\nonumber\\
\ddot{\alpha}_c(0)=0,~~\ddot{\alpha}_c(T)=0.
\label{alfa}
\eeqa
The first line of conditions in Eq. (\ref{alfa}) sets the states at the desired locations. The second one leaves them at rest. 
The third line is not  necessary to achieve commutativity, but it assures the 
continuity of the trap motion. If the second derivatives do not vanish the trap will not be centered 
at $0$ and $d$ according to Eq. (\ref{classical}). This means that instantaneous trap displacements would be required
at the boundary times, for example from $0$ to $\ddot{\alpha}_c(0)/\omega^2$ at time zero. 
Approaching that ideal jump in practice might not be easy. 

In the next section we shall show how to extend these ideas to two ions of different mass. 
Notice that an alternative method described in \cite{Erik1}, the compensating force approach, may formally be applied to ion
chains to avoid excitation. This however requires applying  different forces to ions of different mass, whereas in  the 
available technology in linear Paul traps the forces are proportional to the charge \cite{Mikel2}. 
\section{Dynamical normal modes for two ions in a moving trap}\label{normal mode}
The Hamiltonian for  1D two-ion transport can be written as
\beqa \label{H-origin}
\hat{H}=\frac{\hat{p}_1^2}{2m_1}+\frac{\hat{p}_2^2}{2m_2}+\frac{1}{2}m_1\omega_1^2(\hat{q}_1-Q_0)^2
+\frac{1}{2}m_2\omega_2^2(\hat{q}_2-Q_0)^2+\frac{C_c}{\hat{q}_1-\hat{q}_2},
\eeqa
where $\hat{q}_1$, $\hat{q}_2$, $\hat{p}_1$ and $\hat{p}_2$ are the position and momentum operators of the two ions (we assume the ion 1 to be always on the 
right of ion 2 due to their strong repulsion), 
$C_c=\frac{e^2}{4\pi\epsilon_0}$ is the Coulomb constant  ($\epsilon_0$ the vacuum permittivity), $m_1$ and $m_2$ are the masses of the two ions,  and $\omega_1$ and $\omega_2$ are the (angular) frequencies of the
ions when they move independently in the trap. 
They are related to the spring constant $u_0$ by $u_0=m_1 \omega_1^2=m_2\omega_2^2$.
For equal masses, the Hamiltonian can be separated using center-of-mass and relative coordinates,
see the Appendix A.
Here we focus on different masses,  $m_1\neq m_2$, so the separability does not hold.
An alternative description is given by the dynamical, mass-weighted
normal-mode coordinates for the moving trap
\cite{Mikel2}. In operator form,
\beq
\hat{q}_{\pm}=a_{\pm}\sqrt{m}\left(\hat{q}_1-Q_0-\frac{l}{2}\right)+b_{\pm}\sqrt{\mu m}\left(\hat{q}_2-Q_0+\frac{l}{2}\right),
\eeq
with conjugate momenta
\beq
\hat{P}_{\pm}=\frac{1}{\sqrt{m}}\left(a_{\pm}\hat{p}_1+\frac{b_{\pm}}{\sqrt{\mu}}\hat{p}_2\right),
\eeq
where $l=2\sqrt[3]{\frac{C_c}{4u_0}}$ is the equilibrium distance between the ions, $\mu=m_2/m_1$,
$m_1=m$, and the coefficients
%
\beqa
a_{\pm}&=&\left(\frac{1}{1+(1-\frac{1}{\mu}\mp\sqrt{1-\frac{1}{\mu}+\frac{1}{\mu^2}})^2\mu}\right)^{1/2},
\nonumber\\
b_{\pm}&=&\left(1-\frac{1}{\mu}\mp\sqrt{1-\frac{1}{\mu}+\frac{1}{\mu^2}}\right)\sqrt{\mu}a_{\pm},
\label{12}
\eeqa
are normalized as $a_\pm^2+b_\pm^2=1$. They also obey the orthogonality relation $a_+a_-+b_+b_-=0$, as well as $a_+b_--a_-b_+=1$.  
To write the Hamiltonian for normal mode coordinates, we have to transform 
the original Hamiltonian (\ref{H-origin}) and add the term $\hat{P}_-\dot{\hat{q}}_-+\hat{P}_+\dot{\hat{q}}_+$ since the transformation
of coordinates depends on time through $Q_0(t)$. Note that $\dot{\hat{q}}_\pm$ are just functions of time (``c-numbers''), so they commute with all operators and 
in particular with the momenta. 
Here the classical theory of canonical transformations may be applied and gives the same results than the quantum approach
 in \cite{Mikel2} based on unitary transformations,  
\beqa \label{wholeH}
\hat{H}_{N}&=&\frac{1}{2}(\hat{P}_+-P_{0+})^2+\frac{1}{2}\Omega_+^2\hat{q}^2_+
\nonumber\\
&+&\frac{1}{2}(\hat{P}_--P_{0-})^2+\frac{1}{2}\Omega_-^2\hat{q}^2_-
\nonumber\\
&+&\sum_{n=3}^{\infty}\frac{(-1)^{n}C_c(\hat{q}_1-\hat{q}_2-l)^n}{l^{n+1}},
\eeqa
where we have kept cubic and higher order terms still in the old coordinates for simplicity, and 
\beqa
\Omega^2_{\pm}&=&\omega_1^2\left(1+\frac{1}{\mu}\pm\sqrt{1-\frac{1}{\mu}+\frac{1}{\mu^2}}\right),
\label{ome}
\\
\label{P0}
P_{0\pm}&=&-\dot{\hat{q}}_\pm=\dot{Q}_0\sqrt{m}(a_{\pm}+b_{\pm}\sqrt{\mu}).
\eeqa
The two modes are independent if we neglect cubic and higher order anharmonic and mode-coupling terms.
In the following, we shall use this harmonic approximation to investigate the
effect of spring constant errors and design fast transport protocols to avoid final excitation.
We shall also check the validity of the approximation. 
\section{Spring-constant error}\label{spring constant error}
Let us now consider a modified spring-constant $u_0(1+\lambda)$, where the relative error $\lambda$
with respect to the ideal value $u_0$ remains constant during the transport time.
This implies that, for each ion, the squared frequencies are   $\omega_{1,2}^2(1+\lambda)/(2\pi)$.
The Hamiltonian in the laboratory frame will take the form
\beqa\label{Herror-origin}
\hat{H}&=&\frac{\hat{p}_1^2}{2m_1}+\frac{\hat{p}_2^2}{2m_2}+\frac{1}{2}m_1\omega_1^2(1+\lambda)(\hat{q}_1-Q_0)^2
\nonumber\\
&+&\frac{1}{2}m_2\omega_2^2(1+\lambda)(\hat{q}_2-Q_0)^2+\frac{C_c}{\hat{q}_1-\hat{q}_2}.
\eeqa
We define new coordinates as
\beqa
\hat{q}'_+&=&a_+\sqrt{m}\left(\hat{q}_1-Q_0-\frac{l'}{2}\right)+b_+\sqrt{\mu m}\left(\hat{q}_2-Q_0+\frac{l'}{2}\right),
\nonumber\\
\hat{q}'_-&=&a_-\sqrt{m}\left(\hat{q}_1-Q_0-\frac{l'}{2}\right)+b_-\sqrt{\mu m}\left(\hat{q}_2-Q_0+\frac{l'}{2}\right),
\nonumber\\
\eeqa
where the equilibrium distance is now
\beq
l'=l\left(1+\lambda\right)^{-\frac{1}{3}},
\eeq
whereas the expressions for normal-mode momenta are not affected by the error.
Within the harmonic approximation, the Hamiltonian for normal coordinates becomes
\beqa
\label{Herror-normal}
\hat{H}_{N}&\simeq&\frac{1}{2}(\hat{P}_+-P_{0+})^2+\frac{1}{2}(\hat{P}_--P_{0-})^2
\nonumber\\
&+&\frac{1}{2}\Omega_+^2(1+\lambda)\hat{q}'^2_++\frac{1}{2}\Omega_-^2(1+\lambda)\hat{q}'^2_-.
\eeqa
Now we use a transformation that shifts the momenta to the trap frame, 
$$
\hat{U}_1=e^{-\frac{i}{\hbar}P_{0+}\hat{q}'_+-\frac{i}{\hbar}P_{0-}\hat{q}'_-},
$$
with $[\hat{P}_+,\hat{q}'_+]=[\hat{P}_+,\hat{q}_+]$, $[\hat{P}_-,\hat{q}'_-]=[\hat{P}_-,\hat{q}_-]$.
The corresponding Hamiltonian $\hat{H}'_{N}=\hat{U}_1\hat{H}_{N}\hat{U}_1^{\dag}-i\hbar\hat{U}_1\partial_t\hat{U}_1^{\dag}$
for  the transformed wave function $|\psi'\ra=\hat{U}_1|\psi\ra$ is (neglecting the terms that depend only on time)
\beq\label{H-normal-e}
\hat{H}'_{N}=\hat{H}_++\hat{H}_-,
\eeq
where
\beqa \label{newerrorH}
\hat{H}_+=\frac{\hat{P}_+^2}{2}+\frac{\Omega_+^{'2}}{2}(\hat{q}'_+-q'_{0+})^2,
\nonumber\\
\hat{H}_-=\frac{\hat{P}_-^2}{2}+\frac{\Omega_-^{'2}}{2}(\hat{q}'_--q'_{0-})^2,
\eeqa
with
\beqa
\Omega'^2_{\pm}&=&\Omega_{\pm}^2(1+\lambda),
\nonumber\\
q_{0\pm}'&=&-\dot{P}_{0\pm}/{\Omega'}_{\pm}^2=q_{0\pm}(1+\lambda)^{-1},
\label{auxi}
\nonumber\\
q_{0\pm}&=&-\dot{P}_{0\pm}/\Omega_{\pm}^2.
\eeqa
To have common initial and final states for the dynamics driven by the Hamiltonians (\ref{Herror-normal})  and (\ref{H-normal-e}), and agreement between the Hamiltonians  at these
boundary times,
$P_{0\pm}$ should satisfy the boundary conditions
\beqa
P_{0\pm}(0)&=&P_{0\pm}(T)=0,
\nonumber\\
\dot{P}_{0\pm}(0)&=&\dot{P}_{0\pm}(T)=0,
\eeqa
which implies (from Eq. (\ref{P0}))
\beqa\label{Q0dot}
\dot{Q}_{0}(0)&=&\dot{Q}_{0}(T)=0,
\\
\ddot{Q}_{0}(0)&=&\ddot{Q}_{0}(T)=0.
\label{Q0ddot}
\eeqa
The unperturbed ``trajectories'' $\alpha_{\pm}$ play the role of $\alpha_c$ in each mode; note that they are ``trajectories'' in a normal-mode coordinate space. 
They  satisfy 
\beq\label{alpha-q0}
\ddot{\alpha}_{\pm}+\Omega_{\pm}^2(\alpha_{\pm}-q_{0\pm})=0,
\eeq
as well as the boundary conditions  to make the excitation energy for the 
unperturbed spring constant zero at $T$,  
%
%
\beqa
\label{alpha}\alpha_{\pm}(0)=0,~~~\alpha_{\pm}(T)=0,
\\
\label{alphadot}\dot{\alpha}_{\pm}(0)=0,~~~\dot{\alpha}_{\pm}(T)=0,
\\
\label{alphaddot}\ddot{\alpha}_{\pm}(0)=0,~~~\ddot{\alpha}_{\pm}(T)=0, 
\eeqa
compare them to the ones in Eq. (\ref{alfa}).

The  perturbed trajectories, denoted as  $F_{\pm}(t)$,  satisfy instead
\beq
\ddot{F_{\pm}}(t)+\Omega_{\pm}'^2[F_{\pm}(t)-q'_{0\pm}]=0.
\eeq
Both $\alpha_{\pm}$  and $F_\pm$ may be found by explicit integral expressions.
With or without interaction the functions and their derivatives vanish at  $t=0$,
\beqa
\alpha_\pm(t)&=&{\Omega_\pm}\int_0^t dt'  q_{0\pm}(t') \sin[\Omega_\pm(t-t')],
\\
F_\pm(t)&=&{\Omega'_\pm}\int_0^t dt'  q'_{0\pm}(t') \sin[\Omega'_\pm(t-t')].
\eeqa
The boundary conditions for $F_\pm(0)$ and $\dot{F}_\pm(0)$ may be inferred from the physically motivated  assumption that
the initial state is  the ground state of the Hamiltonian  (\ref{Herror-normal}) irrespective of the $\lambda$ value.

Defining the correction $f_{\pm}(t)$ by $F_{\pm}(t)=\alpha_{\pm}(t)+f_{\pm}(t)$, we have
\beq
\ddot{f_{\pm}}(t)+\Omega_{\pm}'^2f_{\pm}(t)=\lambda\ddot{\alpha}_{\pm}-B_{\pm}\Omega_{\pm}'^2,
\eeq
where
\beq
B_\pm=q_{0\pm}[1-(1+\lambda)^{-1}],
\eeq
which can be solved formally as
\beq
f_{\pm}(t)=
\frac{1}{\Omega_{\pm}'}\int_0^t\bigg[\lambda\ddot{\alpha}_{\pm}(t')-B_{\pm}(t')\Omega_{\pm}'^2\bigg]\sin[\Omega'_{\pm}(t-t')]dt.
\nonumber\\
\eeq

The energy can be calculated exactly within the harmonic approximation and it takes at final time $T$ the form 
\beqa
\la \hat{H}_{\pm}(T)\ra=\la \phi_{\pm}(T;n)|\hat{H}_{\pm}(T)|\phi_{\pm}(T;n)\ra
=\left(n+\frac{1}{2}\right)\hbar\Omega'_{\pm}+E_{e\pm}(T),
\eeqa
where $\phi_{\pm}({q}'_{\pm},T;n)=\exp{\left[i \frac{\dot{F}_{\pm}(T)
{q}'_{\pm}}{\hbar}\right]}\phi^{(0)}[{q}'_{\pm}-F_{\pm}(T);n]$ are the eigenstates of the invariants $\hat{I}_\pm$ corresponding to the Hamiltonians $\hat{H}_{\pm}$ (\ref{newerrorH}), and the final excitation energy for each mode is
\beqa
E_{e\pm}(T)&=&
\frac{1}{2}\left\{\int_0^T\bigg[\lambda\ddot{\alpha}_{\pm}(t)-B_{\pm}(t)\Omega_{\pm}'^2\bigg]\cos(\Omega'_{\pm}t)dt\right\}^2
\nonumber\\
&+&\frac{1}{2}\left\{\int_0^T\bigg[\lambda\ddot{\alpha}_{\pm}(t)-B_{\pm}(t)\Omega_{\pm}'^2\bigg]\sin(\Omega'_{\pm}t)dt\right\}^2.
\nonumber
\label{ee}
\eeqa
%
The total excitation energy is $E_e(T)=E_{e+}(T)+E_{e-}(T)$.

In order to eliminate the excitation energy, the designed protocol should satisfy
\beqa\label{realcondition}
\int_0^T\bigg[\lambda\ddot{\alpha}_{\pm}(t)-B_{\pm}(t)\Omega_{\pm}'^2\bigg]\cos(\Omega'_{\pm}t)dt=0,
\nonumber\\
\int_0^T\bigg[\lambda\ddot{\alpha}_{\pm}(t)-B_{\pm}(t)\Omega_{\pm}'^2\bigg]\sin(\Omega'_{\pm}t)dt=0.
\eeqa
\begin{figure}
\begin{center}
\scalebox{0.65}[0.65]{\includegraphics{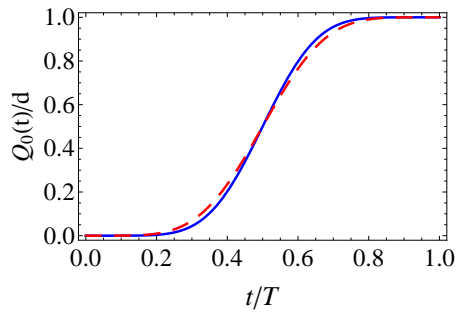}}
\caption{(Color online) Trap trajectories versus time for the protocols
in Eq. (\ref{optimal-poly}) (red dashed line) and Eq. (\ref{cos}) (blue solid line) for $T=10.5$ $T_0$. $\omega_1=2\pi\times2$ MHz, $T_0=2\pi/\omega_1$, $d=370$ $\mu m$, and $n=0$.  The ions (1 and 2) are $^{9}$Be$^+$ and $^{24}$Mg$^+$
respectively.}
\label{figq0}
\end{center}
\end{figure}
Since $\lambda$ is generally not known or drifts from run to run of the experiment, we simplify the condition.
We use the Taylor expansion $(1+\lambda)^{-1}=1-\lambda+\cdot\cdot\cdot$ and keep in the first order of $\lambda$, so
$B_{\pm}\simeq\lambda q_{0\pm}$. We also approximate $\sin(\Omega'_{\pm}t)\simeq\sin(\Omega_{\pm}t)$, $\cos(\Omega'_{\pm}t)\simeq\cos(\Omega_{\pm}t)$, so the dominant order of the condition (\ref{realcondition}), using  Eq. (\ref{alpha-q0}) and dropping constants, is
\beqa\label{error-newcondition}
\int_0^T\alpha_{\pm}(t)\cos(\Omega_{\pm}t)dt=0,
\nonumber\\
\int_0^T\alpha_{\pm}(t)\sin(\Omega_{\pm}t)dt=0.
\eeqa
As $Q_0(t)$ depends on $\alpha_+$ and $\alpha_-$ via Eqs. (\ref{auxi}) and (\ref{P0}),
and it should be a unique function, common to both modes,
we may assume an ansatz for $Q_0(t)$ with free parameters, and then get $\alpha_{\pm}$ from Eq. (\ref{alpha-q0}) to satisfy the conditions (\ref{alpha}), (\ref{alphadot}) and (\ref{alphaddot}).
We  solve Eq. (\ref{alpha-q0}) using the condition (\ref{alpha}). Since Eq. (\ref{Q0ddot}) implies that the condition (\ref{alphaddot}) of $\ddot{\alpha}_{\pm}$ is also satisfied, we should thus design $Q_0$ to satisfy the conditions (\ref{Q0}), (\ref{Q0dot}) and  (\ref{Q0ddot}); and moreover ${\alpha}_{\pm}$ should satisfy Eqs. (\ref{alphadot}),
and the integrals in Eq. (\ref{error-newcondition}) have to be nullified.
In the following, we compare the performance of  different protocols for $Q_0(t)$.

\begin{figure}
\begin{center}
\scalebox{0.65}[0.65]{\includegraphics{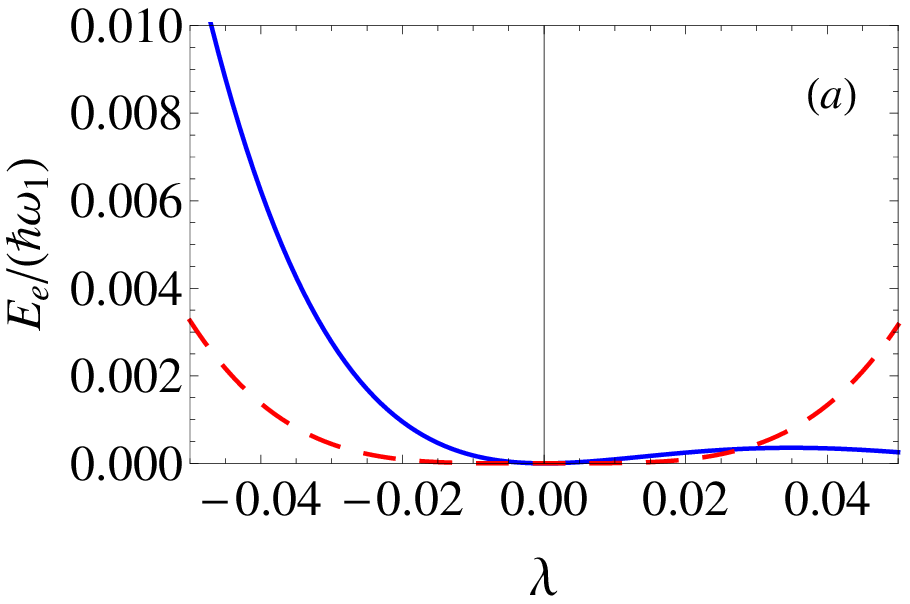}}
\scalebox{0.65}[0.65]{\includegraphics{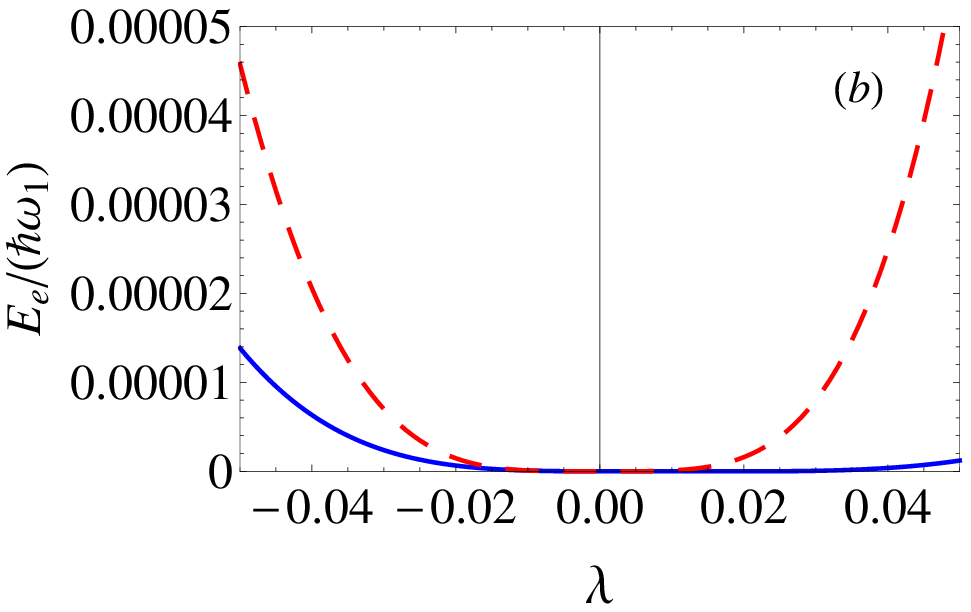}}
\caption{(Color online) Excitation suppression. Excitation energy $E_e$ versus the error for $T=7.5$ $T_0$ (a) and $T=10.5$ $T_0$ (b), 
using the protocol in Eq. (\ref{optimal-poly}) (red dashed line)
and Eq. (\ref{cos}) (blue solid line). Other parameters are the same as in Fig. \ref{figq0}. }
\label{figerror-pop}
\end{center}
\end{figure}

One possible ansatz is the polynomial (see Fig. \ref{figq0})
\beq \label{optimal-poly}
Q_0(t)=\sum_{j=0}^{13}\beta_jt^j,
\eeq
where the coefficients are found from the above $14$ conditions: (\ref{Q0}), (\ref{Q0dot}), (\ref{Q0ddot}), (\ref{alphadot}) and (\ref{error-newcondition}).
The final excitation versus $\lambda$ is shown in Fig. \ref{figerror-pop}
for a couple of final times.

%

\begin{figure}[h]
\begin{center}
\scalebox{0.65}[0.65]{\includegraphics{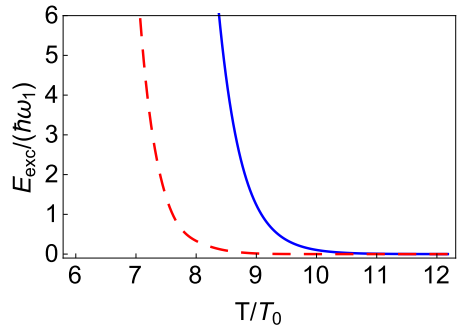}}
\caption{(Color
online) Breakdown of the harmonic approximation at short shuttling times. The excitation energy $E_{exc}$ is depicted versus 
final time for the transport of  the dynamics of two 
classical ions driven by the full Hamiltonian (\ref{H-origin}) using the protocols (\ref{optimal-poly}) (red dashed line) and (\ref{cos}) (blue solid line). The
parameters are the same as in Fig. \ref{figq0} but no harmonic approximation is applied.
The initial  ions are at equilibrium and the final excitation energy $E_{exc}$
is computed as the total classical energy minus the equilibrium energy.}
\label{figclassical}
\end{center}
\end{figure}

%
A simpler option is the trigonometric ansatz, $Q_0(t)=d\big[\beta_0+\sum_{j=1}^4 \beta_j \cos\big(\frac{(2j-1)\pi t}{T}\big)
\big]$, that satisfies the conditions (\ref{Q0}), (\ref{Q0dot}), (\ref{Q0ddot}) and (\ref{alphadot})
with just five parameters,
\beqa \label{cos}
Q_0(t)&=&d\Bigg[\frac{1}{2}+\left(-\frac{9}{16}+2 \beta_3+5 \beta_4\right)\cos\left(\frac{\pi t}{T}\right)
\nonumber
\\
&+&\frac{1}{16}\left(1-48 \beta_3-96\beta_4\right)\cos\left(\frac{3\pi t}{T}\right)
\nonumber\\
&+&\beta_3\cos\left(\frac{5\pi t}{T}\right)+\beta_4\cos\left(\frac{7\pi t}{T}\right)\Bigg],
\eeqa
where $\beta_3$ and $\beta_4$ can be given explicitly as functions of $T$,
\beqa
\beta_3&=&-\frac{49(T^2\Omega_+^2-25\pi^2)(T^2\Omega_-^2-25\pi^2)}{2048T^4\Omega_+^2\Omega_-^2},
\nonumber\\
\beta_4&=&\frac{5(T^2\Omega_+^2-49\pi^2)(T^2\Omega_-^2-49\pi^2)}{2048T^4\Omega_+^2\Omega_-^2}.
\eeqa
\begin{figure}[h]
\begin{center}
\scalebox{0.65}[0.65]{\includegraphics{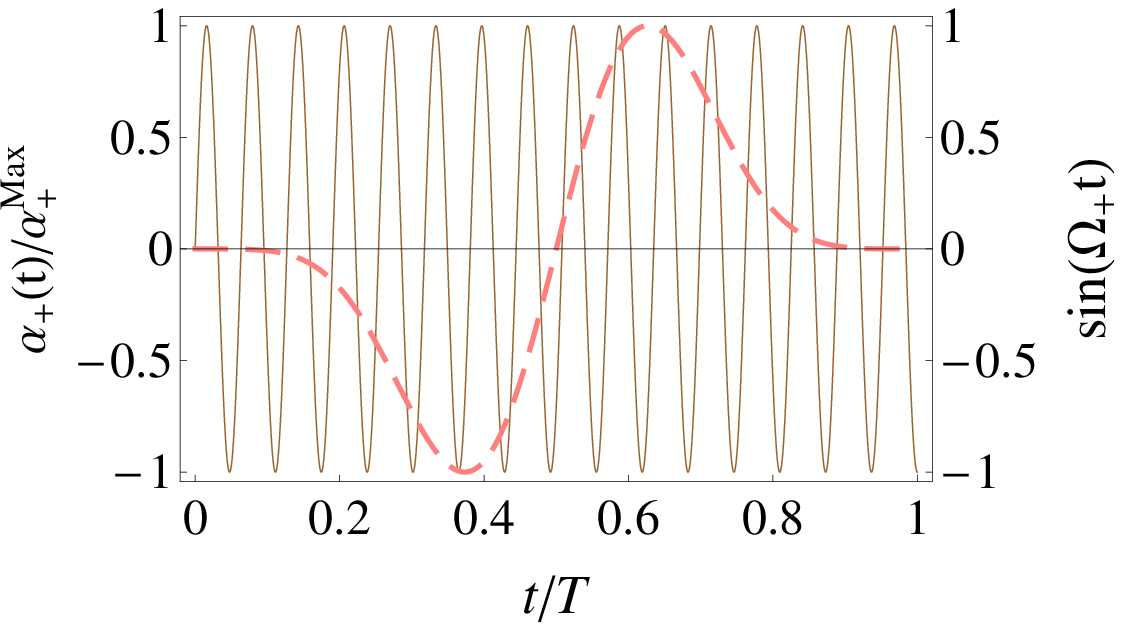}}
\scalebox{0.65}[0.65]{\includegraphics{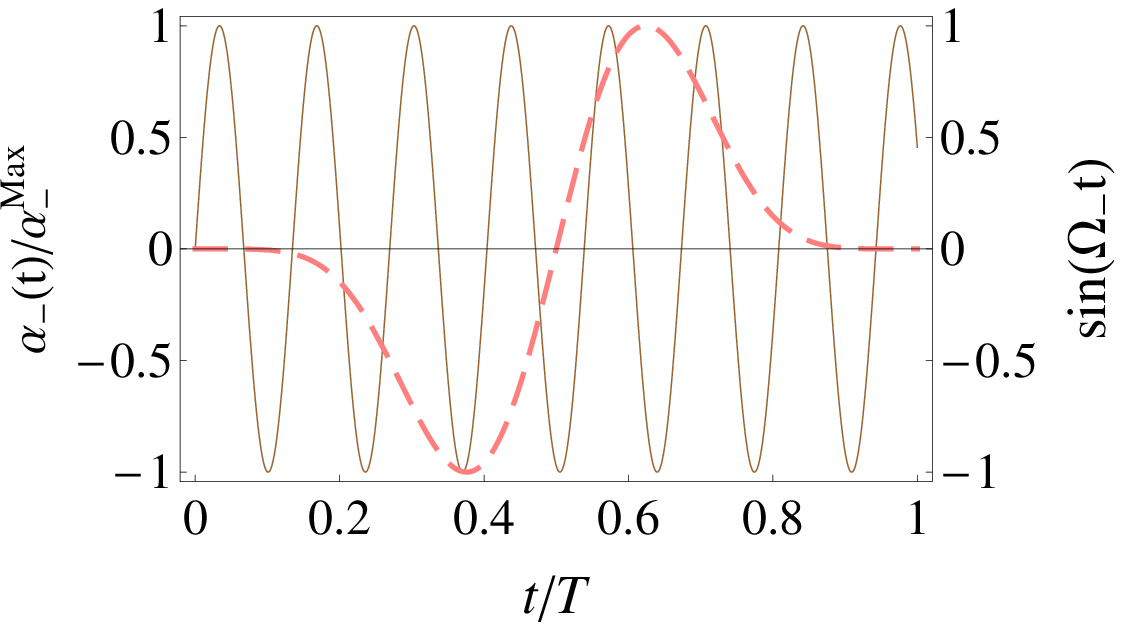}}
\caption{(Color
online) $\alpha_{\pm}$ (normalized, pink dashed line) and $\sin(\Omega_{\pm}t)$ (brown solid line) versus time $t$ for the protocol (\ref{cos}); 
mode + in the left figure, and mode - in the right figure.  $T=10.5$ $T_0$, and other
parameters are the same as in Fig. \ref{figq0}. The rapid oscillation of the sines compared to the slower variation of $\alpha_\pm$ makes the  
the integrals in Eq. (\ref{error-newcondition}) negligible.}
\label{figalpha}
\end{center}
\end{figure}
The behavior of this cosine protocol is quite remarkable. In particular, even though the exact vanishing of the integrals in (\ref{error-newcondition})
is not imposed, they are indeed negligible for
times larger than approximately eight periods (of particle $1$).  (The integrals are doable explicitly 
and the result is given in the Appendix B.) The reason is that the mode trajectories $\alpha_{\pm}$ vary slowly with respect to the faster oscillation of $\sin(\Omega_{\pm}t)$ or $\cos(\Omega_{\pm}t)$,
see Fig. \ref{figalpha}. Of course this cancellation will not hold for very short process times $T$ of the order of a few oscillations, but in that short-time regime the harmonic approximation breaks down anyway. 
As shown in Fig. \ref{figerror-pop}, the cosine protocol (\ref{cos})
is as stable as the polynomial protocol  (\ref{optimal-poly}), even more stable for the longer final time.
The range of validity of the harmonic approximation is examined in
Fig. \ref{figclassical} using classical dynamics. This classical approximation is enough to detect significant deviations from the harmonic behavior
and much less demanding computationally than a full quantum calculation.  
The figure shows the final excitation of two classically moving ions for $\lambda=0$, with respect to the equilibrium energy,  using 
both protocols and the exact Hamiltonian (\ref{H-origin}).
The protocols would be excitation-free for $\lambda=0$ in the harmonic approximation, so the excitation at short times is due to anharmonicities and mode-coupling.
The polynomial ansatz is slightly more robust with respect to them, holding negligible excitation up to nine oscillation periods, versus ten
oscillation periods for the cosines.
%
%
%
The trigonometric ansatz provides in summary an excellent, simple
way to eliminate the spring constant error for the transport of two ions, as it is 
given by explicit time-dependent coefficients.  

\begin{figure}
\begin{center}
\scalebox{0.58}[0.58]{\includegraphics{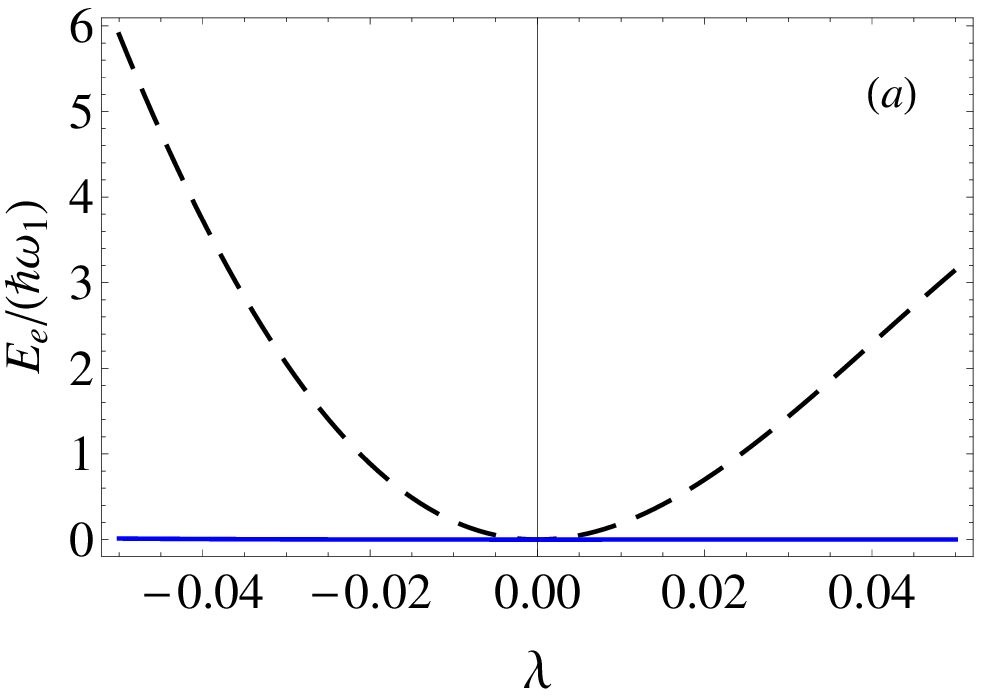}}
\scalebox{0.60}[0.60]{\includegraphics{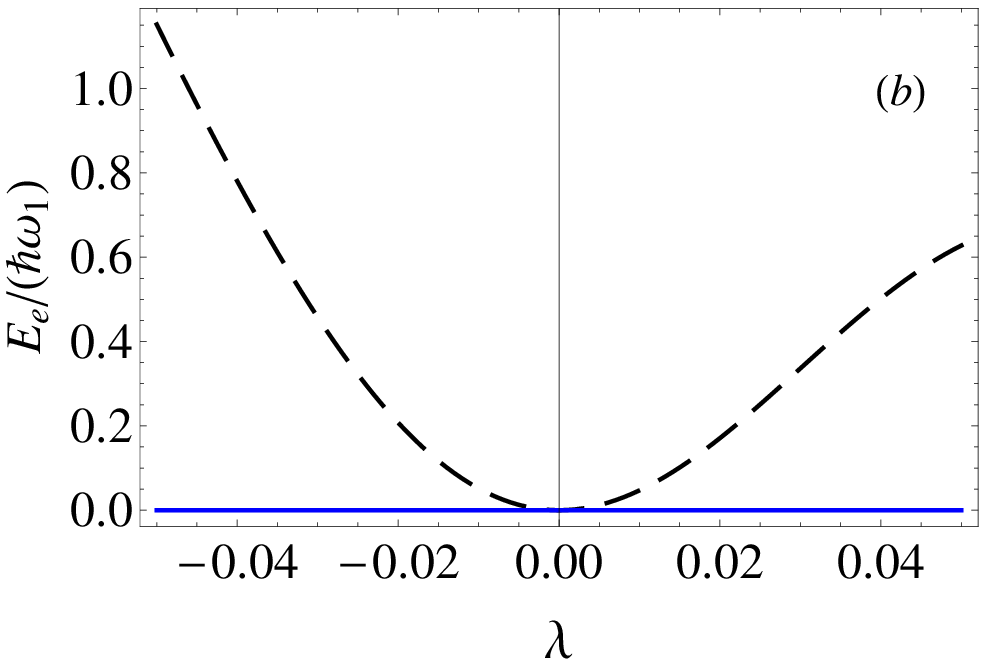}}
\caption{(Color online) Comparison of the excitation energies $E_e$ versus the error for $T=7.5$ $T_0$ (a) and $T=10.5$ $T_0$ (b), using the protocol in 
Eq. (\ref{cos}) (solid blue line), and a protocol where $Q_0$ is designed assuming that the center of mass is uncoupled (dashed black line). Other parameters are the same as in Fig. \ref{figq0}.}
\label{newfig}
\end{center}
\end{figure}

Finally, let us compare the results of the cosine protocol with the protocols derived in \cite{Mainz} for one particle. 
As stated earlier, the CM is coupled to the relative motion unless the masses are equal. For equal masses only the CM 
is relevant to design the trap trajectory. For unequal masses, we may try to engineer a $Q_0(t)$ approximately neglecting the coupling, in other words, 
considering a single uncoupled (CM) particle with Hamiltonian 
$\frac{\hat{P}^2}{2M}+\frac{1}{2}M\omega_{_{CM}}^2(\hat{Q}-Q_0)^2$, 
where $\hat{Q}$ and $\hat{P}$ are conjugate CM position and momentum operators,   
$M=m_1+m_2$, and 
$\omega_{_{CM}}^2=\frac{2u_0}{M}$.  
Specifically we may design $Q_0(t)$ as in \cite{Mainz}, Sec. IV,  to make it robust versus the spring-constant 
errors we are interested in here. With that new $Q_0(t)$ we compute the excitation using Eq. (\ref{ee}). It is indeed much
larger than the excitation of the cosine protocol as shown in Figure \ref{newfig}. We conclude that for unequal masses 
the two-mode approach is clearly superior to a simplistic approach based on a single uncoupled 
CM coordinate.     
%
%
%
\section{Conclusion}
We have found trap trajectories to transport without final excitation two ions of different mass. The trajectories are  designed to be robust
with respect to errors in the spring constant. To achieve that goal we have combined invariant-based inverse engineering and a harmonic approximation in dynamically defined normal-mode coordinates.  Shortcuts to adiabaticity with enhanced robustness have been designed as well
for discrete systems \cite{Andreas, Dijon, Lu}, and some of the results and techniques may be applied to ion transport.
In particular the robustness can be improved systematically if necessary as in \cite{Dijon}, by nullifying integrals associated with higher
orders in the relative error parameter. The design of trap-transport functions  robust with respect to random, noisy perturbations of the spring constant requires a different treatment and will be tackled elsewhere. 
\ack
We are grateful to Margarita and Vladimir Man'ko for their inspiring research throughout the years. 
We dedicate this article to them as a little tribute to their
studies on dynamical invariants.     
This work was supported by the
Grants No. IT472-10, No. FIS2009-12773-C02-01, No. UFI 11/55, No. 11474193, No. 61176118, No. 13PJ1403000, No. 2013310811003,
and the Program for Professor of Special Appointment
(Eastern Scholar) at Shanghai Institutions of Higher Learning. M. P. acknowledges a fellowship by UPV/EHU. 
\appendix
\section{Equal masses}
In this Appendix we discuss the equal mass limit  and explain the connection between
the results for CM and relative coordinates in \cite{Mainz} and the dynamical normal-mode approach followed in this paper.

\textbf{CM-relative coordinates}: In the CM-relative coordinates, the Hamiltonian in (\ref{H-origin}) for $m_1=m_2=m$, can be written as
\beqa
\hat{H}_{}=\frac{\hat{P}^2}{2M}+\frac{1}{2}M\omega_{_{CM}}^2(\hat{Q}-Q_0)^2
+\frac{\hat{p}_r^2}{2m_r}+\frac{1}{2}m_r\omega_r^2\hat{r}^2+\frac{C_c}{\hat{r}},
\label{hamiCMr}
\eeqa
where
\beqa
\hat{Q}&=&\frac{\hat{q}_1+\hat{q}_2}{2}, ~~\hat{r}=\hat{q}_1-\hat{q}_2,
\nonumber\\
\hat{P}&=&\hat{p}_1+\hat{p}_2, ~~\hat{p}_r=\frac{\hat{p}_1-\hat{p}_2}{2},
\nonumber\\
M&=&2m, ~~m_r=\frac{m}{2},~\omega_{_{CM}}=\omega_r=\omega_1.
\eeqa
The two coordinates are uncoupled, and
only the motion of the center mass depends on the trajectory $Q_0(t)$ of the trap.  Thus, the design
of shortcuts without final excitation reduces to an an effective one-particle transport problem.

\textbf{Normal-modes}: For equal masses $\mu=1$, so Eqs. (\ref{12}), (\ref{ome}) and (\ref{P0}) become
\beqa
a_+&=&\frac{1}{\sqrt{2}}, ~~~~~b_+=-\frac{1}{\sqrt{2}},
\nonumber\\
a_-&=&\frac{1}{\sqrt{2}},~~~~~ b_-=\frac{1}{\sqrt{2}},
\\
\Omega_+^2&=&3\omega_1^2,~~~~~\Omega_-^2=\omega_1^2,
\nonumber\\
P_{0+}&=&0,~~~~~~~P_{0-}=\sqrt{2m}\dot{Q}_0.
\eeqa
The new coordinates and momenta are
\beqa
\hat{q}_+&=&\sqrt{\frac{m}{2}}(\hat{q}_1-\hat{q}_2-l),~~
\hat{q}_-=\sqrt{\frac{m}{2}}(\hat{q}_1+\hat{q}_2-2Q_0),
\nonumber\\
\hat{P}_+&=&\frac{1}{\sqrt{2m}}(\hat{p}_1-\hat{p}_2),~~
\hat{P}_-=\frac{1}{\sqrt{2m}}(\hat{p}_1+\hat{p}_2).
\eeqa
Then the Hamiltonian in normal-mode coordinates (\ref{wholeH}) takes the form
\beqa\label{H-equal mass}
\hat{H}_{N}&=&\frac{1}{2}\hat{P}_+^2+\frac{1}{2}\Omega_+^2\hat{q}^2_+
+\sum_{n=3}^{\infty}\left(\frac{2}{m}\right)^{\frac{n}{2}}\frac{(-1)^nC_c\hat{q}_+^n}{l^{n+1}}
\nonumber\\
&+&\frac{1}{2}(\hat{P}_--P_{0-})^2+\frac{1}{2}\Omega_-^2\hat{q}^2_-.
\eeqa
To check the consistency between the normal mode approach and the CM-relative method,
the normal mode coordinates and momenta can be expressed as
\beqa\label{new definition}
\hat{q}_+&=&\sqrt{\frac{m}{2}}(\hat{r}-l),~~
\hat{q}_-=\sqrt{2m}(\hat{Q}-Q_0),
\nonumber\\
\hat{P}_+&=&\sqrt{\frac{2}{m}} \hat{p}_r,~~
\hat{P}_-=\frac{1}{\sqrt{2m}}\hat{P}.
\eeqa
If we apply to the wave function that evolves with $\hat{H}_{N}$ the unitary transformation
\beq
\hat{U}_0=\int d{Q}d{r}d{q}_+d{q}_-|{Q},{r}\ra\la {Q},{r}|{q}_+,{q}_-\ra\la{q}_+,{q}_-|,
\eeq
the new Hamiltonian will be $\hat{H}_{NC}=\hat{U}_0\hat{H}_N\hat{U}_0^{\dag}+i\hbar\frac{\partial{\hat{U}_0}}{\partial t}\hat{U}_0^{\dag}$.
For the first part $\hat{H}_1=\hat{U}_0\hat{H}_N\hat{U}_0^{\dag}$, we substitute the definitions (\ref{new definition}) in the Hamiltonian (\ref{H-equal mass}),
\beqa
\hat{H}_{1}&=&\frac{\hat{p}_r^2}{2m_r}+\frac{1}{2}m_r\omega_r^2\hat{r}^2+\frac{C_c}{\hat{r}}
\nonumber\\
&+&\frac{\hat{P}^2}{2M}+\frac{1}{2}M\omega_{_{CM}}^2(\hat{Q}-Q_0)^2-\dot{Q}_0\hat{P}.
\eeqa
For the second part we calculate
\beqa
\frac{\partial\hat{U}_0}{\partial t}=\int d{Q}d{r}d{q}_+d{q}_-|{Q},{r}\ra\frac{\partial(\delta_Q\delta_r)}{\partial t}\la {q}_+,{q}_-|,
\eeqa
where $\la {Q},{r}|{q}_+,{q}_-\ra=\delta[Q-Q(q_+,q_-)]\delta[r-r(q_+,q_-)]=\delta_Q\delta_r$ and
\beqa
\frac{\partial(\delta_Q\delta_r)}{\partial t}&=&\frac{\partial(\delta_Q)}{\partial t}\delta_r+\delta_Q\frac{\partial(\delta_r)}{\partial t}
=-\dot{Q}_0\delta_Q\delta_r\partial_Q.
\eeqa
The second part is thus
\beq
\hat{H}_2=i\hbar\frac{\partial\hat{U}_0}{\partial t} \hat{U}_0^\dagger=\dot{Q}_0\hat{P},
\eeq
so the Hamiltonian $\hat{H}_{NC}=\hat{H}_1+\hat{H}_2$ coincides with
the one in Eq. (\ref{hamiCMr}).
\section{Integrals in the cosine protocol}
Using the protocol (\ref{cos}), we get
\beqa
&&\Omega_\pm^2\int_0^T\alpha_\pm(t)\cos(\Omega_{\pm}t)dt+\Omega_\pm^2\int_0^T\alpha_\pm(t)\sin(\Omega_{\pm}t)dt
\nonumber\\
&&=\frac{11025d\pi^8c_\pm\Omega_\pm(\Omega_\mp^2-\Omega_\pm^2)}
{2D_+}
[1+\cos(\Omega_\pm T)-\sin(\Omega_\pm T)],
\nonumber
\eeqa
where
$
D_\pm=(11025\pi^8-12916\pi^6T^2\Omega_\pm^2+1974\pi^4T^4\Omega_\pm^4-84\pi^2T^6\Omega_\pm^6+T^8\Omega_\pm^8)\Omega_\mp^2.
$
One of the integrals is  zero when $\Omega_\pm T=2k\pi+\frac{\pi}{2}$ or $2k\pi+\pi$, with $k=0,1,2,...$.


\section*{References}
\begin{thebibliography}{999}

\bibitem{Wineland2002} Kielpinski D, Monroe C and  Wineland D 2002 \emph{Nature (London)} \textbf{417}, 709
\bibitem{Rowe} Rowe M A, Ben-Kish A, Demarco B, Leibfried D, Meyer V, Beall J, Britton J, Hughes J, Itano W M, Jelenkovi\'{c} B,  Langer C, Rosenband T and Wineland D J 2002 \emph{Quant. Inf. Comput.} \textbf{2}, 257
%
\bibitem{Wineland2006} Reichle R, Leibfried D, Blakestad R B, Britton J, Jost J D, Knill E, Langer C, Ozeri R, Seidelin S and Wineland D J 2006 \emph{Fortschr. Phys.} \textbf{54}, 666
%
\bibitem{Roos} Roos C 2012 \emph{Physics} \textbf{5}, 94
%
\bibitem{Monroe} Monroe C and Kim J 2013 \emph{Science} \textbf{339}, 1164

\bibitem{2dif}Barrett M D et al. 2003 \emph{Phys. Rev. A} \textbf{68}, 042302
\bibitem{Bowler} Bowler R, Gaebler J, Lin Y, Tan T R, Hanneke D, Jost J D, Home J P, Leibfried D and Wineland D J 2012 \emph{Phys. Rev. Lett.} \textbf{109}, 080502
\bibitem{Schmidt} Walther A, Ziesel F, Ruster T, Dawkins S T, Ott K, Hettrich M, Singer K, Schmidt-Kaler F and Poschinger U 2012 \emph{Phys. Rev. Lett.} \textbf{109}, 080501

\bibitem{Mikel} Palmero M, Torrontegui E, Gu\'{e}ry-Odelin D and Muga J G 2013 \emph{Phys. Rev. A} \textbf{88}, 053423
\bibitem{Mikel2} Palmero M, Bowler R, Gaebler J P, Leibfried D and Muga J G 2014  \emph{Phys. Rev. A} \textbf{90}, 053408 


\bibitem{Mainz}Lu X J, Muga J G, Chen X, Poschinger U G, Schmidt-Kaler F and Ruschhaupt A 2014 \emph{Phys. Rev. A} \textbf{89}, 063414
\bibitem{Erik1} Torrontegui E, Ibanez S, Chen X, Ruschhaupt A, Gu\'{e}ry-Odelin D and Muga J G 2011 \emph{Phys. Rev. A} \textbf{83}, 013415
\bibitem{ManDodo} Dodonov V V and Manko V I, Theory of Nonclassical States
of Light (CRC Press, London, 2003)

\bibitem{other} Harari G, Ben-Aryeh Y and Mann A, arXiv: 1305.2590
\bibitem{LR}Lewis H R and Riesenfeld W B 1969 \emph{J. Math. Phys.} \textbf{10}, 1458
\bibitem{LL} Lewis H R and Leach P G 1982 \emph{J. Math. Phys. }\textbf{23}, 2371
\bibitem{DL} Dhara A K and Lawande S W 1984 \emph{J. Phys. A} \textbf{17}, 2324



\bibitem{Andreas} Ruschhaupt A, Chen X, Alonso D and Muga J G 2012
\emph{New J. Phys.} \textbf{14} 093040
\bibitem{Dijon} Daems D, Ruschhaupt A, Sugny D and Guerin S 2013
\emph{Phys. Rev. Lett.} \textbf{111} 050404

\bibitem{Lu} Lu X J, Chen X, Ruschhaupt A, Alonso D, Guerin S and Muga J G 2013
\emph{Phys. Rev. A} \textbf{88} 033406


\end{thebibliography}
\end{document}